\documentstyle[pre,aps,twocolumn]{revtex}

\begin{document} 
\draft 
\title{Semiclassical Non-Trace Type Formulas for Matrix Element
Fluctuations and Weighted Densities of States}
\author{J\"org Main$^1$ and G\"unter Wunner$^2$}
\address{$^1$Institut f\"ur Theoretische Physik I,
         Ruhr-Universit\"at Bochum, D-44780 Bochum, Germany}
\address{$^2$Institut f\"ur Theoretische Physik und Synergetik,
         Universit\"at Stuttgart, D-70550 Stuttgart, Germany}

\date{April 1, 1999}
\maketitle

\begin{abstract}
Densities of states weighted with the diagonal matrix elements of two
operators $\hat A$ and $\hat B$, i.e., $\rho^{(A,B)}(E)
=\sum_n\langle n|\hat A|n\rangle\langle n|\hat B|n\rangle\delta(E-E_n)$
cannot, in general, be written as a trace formula, and therefore no simple
extension of semiclassical trace formulas is known for this case.
However, from the high resolution analysis of quantum spectra in the
semiclassical regime we find strong evidence that weighting the 
$\delta$-functions in the quantum mechanical density of states with the 
product of diagonal matrix elements, 
$\langle n|\hat A|n\rangle\langle n|\hat B|n\rangle$, is equivalent 
to weighting the periodic orbit contributions in the semiclassical 
periodic orbit sum with the product of the periodic orbit means, 
$\langle A\rangle_p\langle B\rangle_p$, of the classical observables 
$A$ and $B$.
Results are presented for the hydrogen atom in a magnetic field for both
the chaotic and near-integrable regime, and for the circle billiard.

\end{abstract}


\section{Introduction}
Semiclassical trace formulas for both chaotic \cite{Gut67,Gut90} and regular 
\cite{Ber76} systems relate quantum spectra and classical periodic orbits.
These formulas have proven to be useful in the analysis of level statistics
\cite{Ber85} and long-range correlations \cite{Win87} in the quantum spectra, 
and it has even become possible to compute individual eigenenergies from 
these expressions \cite{Cvi89,Sie91,Ber92,Mai97a,Mai98}.
Gutzwiller's trace formula \cite{Gut67,Gut90} and the Berry-Tabor formula
\cite{Ber76} are semiclassical approximations to the density of states but
do not provide information about experimentally measurable observables, i.e.,
matrix elements of Hermitian operators.
The trace formulas have been extended to the calculation of diagonal matrix
elements of smooth operators in Refs.\ \cite{Wil88,Eck92}.
The extended trace formulas relate the diagonal matrix elements of operators
to the periodic orbit means of the corresponding classical observables.

However, these formulas cannot be applied directly for the semiclassical
calculation of {\em products} of diagonal matrix element where the 
weighted density of states cannot, in general, be written as a trace formula.
Products of diagonal matrix elements are important in several interesting 
applications of semiclassical theories, e.g., for the semiclassical theory 
of matrix element fluctuations \cite{Eck95}, with the variance of an operator 
$\hat A$ in an eigenstate $|n\rangle$ given by 
${\rm Var}_n A\equiv\langle n|\hat A^2|n\rangle-\langle n|\hat A|n\rangle^2$.
A semiclassical periodic orbit formula for products of diagonal matrix 
elements is also of crucial importance for the semiclassical quantization 
technique developed in Ref.\ \cite{Mai99}, where the classical information 
of a set of observables is used to significantly improve the convergence 
properties of periodic orbit quantization.

In this paper we investigate {\em non-trace} type formulas for the density
of states weighted with the diagonal matrix elements of two operators 
$\hat A$ and $\hat B$, i.e., $\rho^{(A,B)}(E)
=\sum_n\langle n|\hat A|n\rangle\langle n|\hat B|n\rangle\delta(E-E_n)$.
From the high resolution analysis of quantum spectra in the semiclassical 
regime we find strong evidence that weighting the $\delta$-functions in the 
quantum mechanical density of states with the product of diagonal matrix 
elements, $\langle n|\hat A|n\rangle\langle n|\hat B|n\rangle$, is equivalent 
to weighting the periodic orbit contributions in the semiclassical periodic 
orbit sum with the product of the periodic orbit means, 
$\langle A\rangle_p\langle B\rangle_p$, of the classical observables 
$A$ and $B$.

The outline of the paper is as follows.
In Sec.\ \ref{trace:sec} we first briefly review Gutzwiller's trace formula 
for chaotic systems and the Berry-Tabor formula for integrable systems, and 
discuss the extension of both equations to the calculation of diagonal 
matrix elements.
We then apply the theories to systems with scaling properties, and introduce
the high resolution analysis (harmonic inversion) of quantum spectra as
a powerful tool to numerically verify the validity of the semiclassical 
expressions.
In Sec.\ \ref{non-trace:sec} we present our results on the semiclassical
non-trace type formulas.
Strong numerical evidence for the validity of the non-trace type equations
is provided by the harmonic inversion of spectra of two different systems, 
viz.\ the hydrogen atom in a magnetic field and the circle billiard.
Sec.\ \ref{conclusion} concludes with remarks on useful and important 
applications, and an outlook on possible generalizations of the non-trace 
type formulas.

\section{Semiclassical trace formulas}
\label{trace:sec}
\subsection{Matrix element extension of periodic orbit theory}
The quantum mechanical density of states can be written as the trace of 
the Green's function, 
$\rho(E)=\sum_n\delta(E-E_n)=-(1/\pi)\,{\rm Im}\,{\rm tr}\,\hat G_E^+$.
Replacing the quantum mechanical Green's function,
$\hat G_E^+=(E-\hat H+i\epsilon)^{-1}$ with its semiclassical analogue 
and calculating integrals and traces in stationary phase approximation 
Gutzwiller derived the fundamental equation of {\em periodic orbit theory} 
\cite{Gut67,Gut90}, i.e., the density of states expressed in terms of 
quantities of the periodic orbits of the classical system.
To obtain the density of states weighted with the diagonal matrix elements 
of an operator $\hat A$ we start from the generalized trace formula
\begin{eqnarray}
     \rho^{(A)}(E)
 &=& -{1\over\pi}\, {\rm Im}\, {\rm tr} \left(\hat G_E^+\hat A\right)
     \nonumber \\
 &=& -{1\over\pi}\, \lim_{\epsilon\to 0} {\rm Im}\, \sum_n
        {\langle n|\hat A|n\rangle\over E-E_n+i\epsilon}
     \nonumber \\
 &=& \sum_n \langle n|\hat A|n\rangle \delta(E-E_n) \; .
\label{rho_A_qm:eq}
\end{eqnarray}
The r.h.s.\ of Eq.\ \ref{rho_A_qm:eq} is the density of states weighted with 
the diagonal matrix elements $\langle n|\hat A|n\rangle$ of the operator 
$\hat A$.
The semiclassical approximation to Eq.\ \ref{rho_A_qm:eq} for a system with
$N$ degrees of freedom reads \cite{Wil88,Eck92}
\begin{eqnarray}
\label{rho_A_sc:eq}
 & & \rho^{(A)}(E) = \rho^{(A)}_0(E) \\
 &+& {1\over\pi\hbar}\, {\rm Re}\, \sum_p A_p
     \sum_{r=1}^\infty {T_p\over\sqrt{|\det(M_p^r-I)|}}\, 
     e^{i[S_p(E)/\hbar-{\pi\over 2}\mu_p]r} \; , \nonumber
\end{eqnarray}
where the Weyl term
$\rho^{(A)}_0(E)=h^{-N} \int d{\bf q}d{\bf p} A({\bf q},{\bf p})
\delta(E-H({\bf q},{\bf p}))$
is a smooth function of the energy and the fluctuating part is given by the 
periodic orbit sum, with $T_p$ the time period, $S_p$ the classical action, 
$M_p$ the monodromy matrix, and $\mu_p$ the Maslov index of the primitive 
periodic orbit $p$.
The integer $r$ is the repetition number of the orbit.
The weights $A_p$ in the periodic orbit sum (\ref{rho_A_sc:eq}) are the means
of the observable $A$ along the periodic orbit $p$, i.e.
\begin{equation}
 A_p = {1\over T_p} \int_0^{T_p} A({\bf q}_p(t),{\bf p}_p(t)) dt \; .
\label{Ap_t:eq}
\end{equation}
The derivation of Eq.\ \ref{rho_A_sc:eq} requires smoothness of the
observable $A$ over regions in phase space of size $h^N$ \cite{Eck92}.
A rigorous mathematical proof of the semiclassical trace formula 
(\ref{rho_A_sc:eq}) using a coherent states decomposition can be found 
in \cite{Com97}.

In Refs.\ \cite{Wil88,Eck92,Com97} formulas for the semiclassical 
calculation of diagonal matrix elements are obtained for chaotic systems
with isolated periodic orbits.
For {\em regular} systems the semiclassical trace formula for the density 
of states has been derived by Berry and Tabor \cite{Ber76}.
For simplicity we restrict ourselves to systems with two degrees of freedom.
Assuming now that the Hamiltonian is classically integrable, one can 
express it in action-angle variables $({\bf I},\varphi)$ with 
$\varphi_1,\varphi_2\in [0,2\pi]$ as $H({\bf I})$.
For a given torus, $\omega_i=\partial H/\partial I_i$ $(i=1,2)$ are the
corresponding angular frequencies.
Periodic orbits are associated with tori such that the rotation number
$\alpha\equiv \omega_1/\omega_2$ is rational, i.e., $\alpha=M_1/M_2$ with
$M_1$ and $M_2$ integers.
The fluctuating part of the Berry-Tabor formula reads
\begin{eqnarray}
 & & \rho_{\rm fl}(E) \nonumber \\
 &=& {1\over\pi\hbar^{3/2}}\, {\rm Re}\, \sum_{\bf M}
 {T_{\bf M}\over M_2^{3/2}|g_E''|^{1/2}} \,
 e^{i\left(S_{\bf M}(E)/\hbar-{\pi\over 2}\eta_{\bf M}-{\pi\over 4}\right)}\; ,
\label{BT:eq}
\end{eqnarray}
with ${\bf M}=(M_1,M_2)$ specifying the periodic orbit, and $T_{\bf M}$,
$S_{\bf M}$, and $\eta_{\bf M}$ the time, action and Maslov index of the
orbit, respectively.
The function $g_E$ in (\ref{BT:eq}) is obtained by inverting the Hamiltonian, 
expressed in terms of the actions $(I_1,I_2)$ of the corresponding torus,
with respect to $I_2$, viz.\ $H(I_1,I_2=g_E(I_1))=E$ \cite{Boh93}.
By analogy with Eq.\ \ref{rho_A_sc:eq} for chaotic systems the Berry-Tabor 
formula (\ref{BT:eq}) can now be generalized straightforwardly to the 
semiclassical calculation of diagonal matrix elements \cite{Meh99}, yielding
\begin{eqnarray}
\label{BT_A:eq}
 & & \rho^{(A)}(E)= \rho^{(A)}_0(E) \\
 &+& {1\over\pi\hbar^{3/2}}\, {\rm Re}\,
     \sum_{\bf M} A_{\bf M} {T_{\bf M}\over M_2^{3/2}|g_E''|^{1/2}} \,
     e^{i[S_{\bf M}(E)/\hbar-{\pi\over 2}\eta_{\bf M}-{\pi\over 4}]}\, ,
 \nonumber
\end{eqnarray}
with
\begin{equation}
 A_{\bf M} = {1\over(2\pi)^2}\int_0^{2\pi}d\varphi_1\int_0^{2\pi}d\varphi_2
  A(I_1,I_2,\varphi_1,\varphi_2)
\label{Ap_reg:eq}
\end{equation}
the classical average of the observable $A$ on the torus.

\subsection{Scaling systems}
In the following we will apply Eqs.\ \ref{rho_A_sc:eq} and \ref{BT_A:eq}
to systems with scaling properties.
In scaling systems the classical phase space structure does not change
for all values of an appropriate scaling parameter, $w$.
The scaling parameter is usually some power of an external field strength 
or, for Hamiltonians with homogeneous potentials, the energy.
Examples are billiard systems \cite{Meh99} or atoms in magnetic fields 
\cite{Fri89,Has89,Wat93}.
In scaling systems the shape of periodic orbits does not depend on the 
scaling parameter, $w$, and the classical action $S_p$ scales as 
\begin{equation}
 S_p = w s_p \; .
\label{S_po}
\end{equation}
The scaling parameter plays the role of an inverse effective Planck constant,
i.e., $w\equiv\hbar_{\rm eff}^{-1}$.
For scaling systems the weighted densities of states, Eqs.\ \ref{rho_A_sc:eq}
and \ref{BT_A:eq} can be rewritten as a function of the scaling parameter $w$,
i.e.
\begin{eqnarray}
  &&   \rho^{(A)}(w)
  =  \rho^{(A)}_0(w) \nonumber \\
 &+& {1\over\pi\hbar}\, {\rm Re}\, \sum_p A_p \sum_{r=1}^\infty
     {s_p\over\sqrt{|\det(M_p^r-I)|}}\, e^{i[s_p w-{\pi\over 2}\mu_p]r}
\label{rho_A_scaled:eq}
\end{eqnarray}
for chaotic systems, and
\begin{eqnarray}
 &&\rho^{(A)}(w) = \rho^{(A)}_0(w) \nonumber \\
 &+& {1\over\pi\hbar^{3/2}}\, {\rm Re}\, \sum_{\bf M} A_{\bf M}
 {s_{\bf M}\over M_2^{3/2}|g_E''|^{1/2}} \,
 e^{i[s_{\bf M}w-{\pi\over 2}\eta_{\bf M}-{\pi\over 4}]}
\label{BT_A_scaled:eq}
\end{eqnarray}
for two-dimensional systems with regular dynamics.
Note that the time periods $T_p$ and $T_{\bf M}$ in Eqs.\ \ref{rho_A_sc:eq}
and \ref{BT_A:eq} must be replaced with the scaled actions $s_p$ and 
$s_{\bf M}$.
Furthermore the time average of the classical observable $A$ 
(Eq.\ \ref{Ap_t:eq}) must be replaced with the average with respect to the
scaled action,
\begin{equation}
 A_p = {1\over s_p} \int_0^{s_p} A({\bf q}_p(s),{\bf p}_p(s)) ds \; .
\label{Ap_s:eq}
\end{equation}
If an observable $A$ is chosen which is invariant under the scaling of 
the system [or scales $\sim w^\beta$ with a constant exponent $\beta$]
the periodic orbit amplitudes and scaled actions in Eqs.\
\ref{rho_A_scaled:eq} and \ref{BT_A_scaled:eq} do not depend on $w$
[despite a possible power law scaling of the amplitudes with $w^\beta$
which can be transfered to the l.h.s.\ of Eqs. \ref{rho_A_scaled:eq} 
and \ref{BT_A_scaled:eq}].
The attractive feature of scaling systems is that the semiclassical 
weighted density of states [or more generally the density of states 
multiplied by $w^\beta$] is a superposition of sinusoidal functions of 
the scaling parameter $w$.
The Fourier transforms of $w^\beta\rho^{(A)}(w)$ should therefore exhibit 
sharp peaks at the positions of the scaled actions of the periodic orbits.
When analyzing quantum spectra, we will make use of the scaling advantages 
in the following.

\subsection{Precision check of the semiclassical trace formulas}
We now wish to apply the semiclassical trace formulas, Eqs.\ 
\ref{rho_A_scaled:eq} and \ref{BT_A_scaled:eq}, to a physical system with 
chaotic and regular dynamics, respectively, and to check numerically the
validity of the semiclassical equations.
The numerical check is not motivated by doubts on the validity of these 
expressions, which have been mathematically proven, rather we want to 
introduce a powerful numerical technique for the high precision check of 
equations of this kind.
We will demonstrate the accuracy of the method on the well established 
semiclassical trace formulas here and then apply the same technique to 
numerically verify our conjecture on semiclassical non-trace type formulas 
in Sec.\ \ref{non-trace:sec}.

The semiclassical trace formulas can be tested, in principle, by the 
Fourier transform analysis of quantum spectra.
The Fourier transformed spectra should exhibit peaks at the periods
(scaled actions) of periodic orbits with amplitudes given by the
semiclassical expressions.
However, the transformation of spectra with finite length yields limited
resolution only, due to the uncertainty principle of the Fourier transform,
which implies a fundamental restriction to high precision checks of the 
semiclassical trace formulas.
We therefore adopt the method of Ref.\ \cite{Mai97b} where we introduced
{\em harmonic inversion} as a high resolution method for the analysis of
quantum spectra.
We briefly review the basic ideas of the harmonic inversion technique
and refer the reader to Ref.\ \cite{Mai97b} for more details.

According to Eqs.\ \ref{rho_A_scaled:eq} and \ref{BT_A_scaled:eq} the 
semiclassical weighted density of states can be written as the sum of a 
smooth background $\rho^{(A)}_0(w)$ and oscillatory modulations induced 
by the periodic orbits, 
\begin{equation}
\label{semicl}
 \rho^{(A)}(w) = \rho^{(A)}_0(w) + {\rm Re} \sum_p d^{(A)}_p e^{i s_p w} \; .
\end{equation}
The amplitudes $d^{(A)}_p$ and scaled actions $s_p$ of the periodic orbits
are obtained from classical calculations and are in general complex 
quantities.
The amplitudes $d^{(A)}_p$ contain the phase information determined by the 
Maslov indices of orbits and the classical means of the observable $A$
given by Eqs.\ \ref{Ap_reg:eq} and \ref{Ap_s:eq} for regular and chaotic 
systems, respectively.
Instead of using the standard Fourier analysis to extract the amplitudes 
and actions, we adjust a finite range of the quantum spectrum by the 
semiclassical expression (\ref{semicl}) with unknown and in general complex 
parameters $d^{(A)}_p$ and $s_p$.
The problem of fitting a ``signal'' $\rho^{(A)}(w)$ to the functional form
(\ref{semicl}) is known as harmonic inversion.
As a numerical technique for the harmonic inversion of a signal, i.e.\ a 
quantum spectrum, we apply the method of filter-diagonalization 
\cite{Wal95,Man97} which allows extracting the spectral quantities in any 
given interval of interest.
Operationally, one proceeds by setting up a small generalized eigenvalue 
problem.
The actions $s_p$ in the chosen spectral domain and amplitudes 
$d^{(A)}_p$ are obtained from the resulting eigenvalues and eigenvectors.
Thus, the recurrence spectrum is effectively discretized, the number
of terms being the number of eigenvalues in the spectral domain.
This method is a variational one (as opposed to the Fourier 
transform) and therefore practically has an infinite resolution
once the amount of information contained in the signal $\rho^{(A)}(w)$
is greater than the total number of unknowns $d^{(A)}_p$ and $s_p$.

As a physical system for the high precision analysis of quantum spectra
and the comparison with the semiclassical trace formulas we choose the
hydrogen atom in a magnetic field \cite{Fri89,Has89,Wat93}.
This is a scaling system, with $w=\gamma^{-1/3}=\hbar_{\rm eff}^{-1}$ the 
scaling parameter and $\gamma=B/(2.35\times 10^5\, {\rm T})$ the magnetic 
field strength in atomic units.
Introducing scaled coordinates $\gamma^{2/3}{\bf r}$ and momenta 
$\gamma^{-1/3}{\bf p}$ and choosing the projection of the angular momentum 
on the magnetic field axis $L_z=0$ one arrives at the scaled Hamiltonian 
\begin{equation}
   \tilde H = {1\over 2}{\bf p}^2 - {1\over r} + {1\over 8}(x^2+y^2)
 = \tilde E \; ,
\label{H_scal:eq}
\end{equation}
with $\tilde E=E\gamma^{-2/3}$ the scaled energy.
The classical dynamics is near-integrable at low energies, $\tilde E<-0.5$,
and undergoes a transition from regularity to chaos in the energy region
$-0.5<\tilde E<-0.13$.
At energies above $\tilde E=-0.13$ a Poincar\'e surface of section analysis
of the classical dynamics does not exhibit any regular structures larger 
than of microscopic size \cite{Has89}.
We compare spectra at constant scaled energy $\tilde E=-0.1$ with the
results of the semiclassical trace formula (\ref{rho_A_scaled:eq}) for
chaotic systems, and spectra in the near-integrable regime at $\tilde E=-0.5$
with the extended Berry-Tabor formula (\ref{BT_A_scaled:eq}).
We choose two different operators.
The first,
\begin{equation}
 \hat A = {1\over 2r{\bf p}^2}
\label{Adef:eq}
\end{equation}
has already served to study the distribution of transition matrix elements 
in classically chaotic and mixed quantum systems \cite{Boo95,Boo96}.
The second operator is
\begin{equation}
 \hat B = r{\bf p}^2 \; .
\label{Bdef:eq}
\end{equation}
Eigenvalues of the scaling parameter $w$ are obtained by solving
Schr\"odinger's equation (in semiparabolic coordinates
$\mu=\sqrt{r+z}$ and $\nu=\sqrt{r-z}$)
\begin{eqnarray}
 & & \left[2\tilde E ({ \mu}^2+{ \nu}^2)
   - {1\over 4}\mu^2\nu^2(\mu^2+\nu^2) + 4 \right] \, \Psi(\mu,\nu) \nonumber\\
 &=& w^{-2} \left(\hat p_\mu^2 + \hat p_\nu^2\right) \, \Psi(\mu,\nu)  \; ,
\label{H_eps}
\end{eqnarray}
with the radial operators ${{\hat { p}}_\mu}^2$ and 
${{\hat { p}}_\nu}^2$ defined as
\[
   \hat p_\mu^2
 = -{1\over\mu}{\partial \over {\partial\mu}}
   \left( \mu{\partial \over {\partial\mu}}\right) \; , \quad
   \hat p_\nu^2
 = -{1\over\nu}{\partial \over {\partial\nu}}
   \left( \nu{\partial \over {\partial\nu}}\right) \; .
\]
Eq.\ \ref{H_eps} can be written in matrix form by using an appropriate
basis set.
The resulting generalized eigenvalue problem is solved numerically.
It has to be noted that the eigenvectors obtained, $|\psi_n\rangle$, 
are orthonormal with respect to the scaled momentum operator, i.e.
\begin{equation}
   \langle\psi_m|\hat p_\mu^2 + \hat p_\nu^2|\psi_n\rangle
 = \langle m|n\rangle = \delta_{mn} \; ,
\end{equation}
with modified eigenvectors $|n\rangle$ defined by $|n\rangle\equiv
(\hat p_\mu^2+\hat p_\nu^2)^{1/2}|\psi_n\rangle$.
The diagonal matrix elements of an operator $\hat A$ are therefore obtained as
\begin{equation}
 A_{nn} = \langle n|\hat A|n \rangle
 = \langle\psi_n|\hat A (\hat p_\mu^2 + \hat p_\nu^2)|\psi_n \rangle \; .
\end{equation}
We are now prepared to compare the quantum spectra of the hydrogen atom in
a magnetic field with the semiclassical approximations in the chaotic and
regular regime of the classical phase space.

\subsubsection{Chaotic regime}
\begin{figure}
\vspace{9.5cm}
\includegraphics{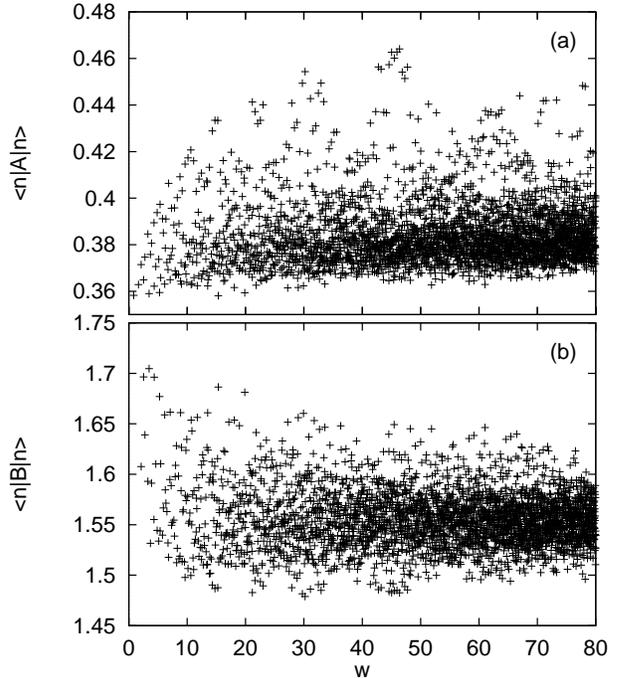}
\caption{\label{fig1} 
Values of the diagonal matrix elements $\langle n|\hat A|n\rangle$ for the 
hydrogen atom in a magnetic field at scaled energy $\tilde E=-0.1$ in the 
chaotic region of phase space as functions of the scaling parameter
$w=\gamma^{-1/3}$ ($\gamma\sim$ magnetic field strength):
(a) operator $\hat A=1/(2r{\bf p}^2)$; (b) $\hat B = r{\bf p}^2$.
}
\end{figure}

\begin{figure}
\vspace{9.7cm}
\includegraphics{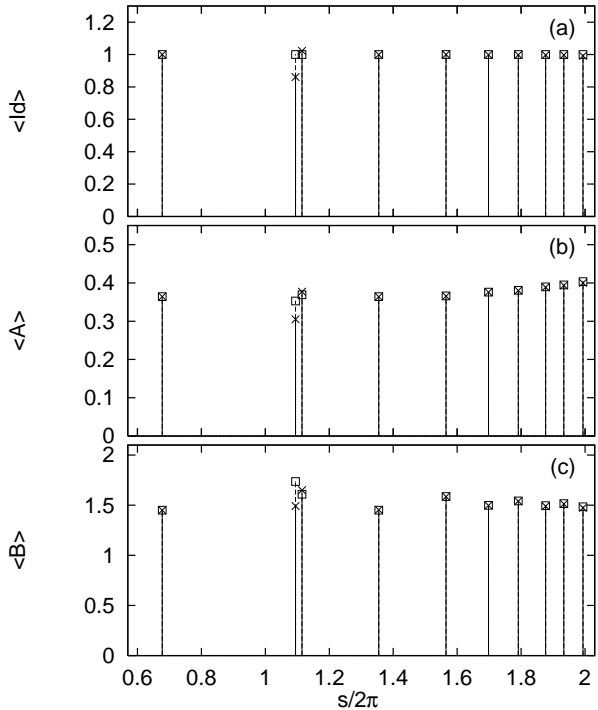}
\caption{\label{fig2} 
Periodic orbit means of observables (a) the identity, (b) $A=1/(2r{\bf p}^2)$,
and (c) $B = r{\bf p}^2$ for the hydrogen atom in a magnetic field at scaled
energy $\tilde E=-0.1$ as functions of the scaled action.
Solid lines and crosses: Results of the harmonic inversion of quantum spectra.
Dashed lines and squares: Periodic orbit means obtained by classical
calculations.
The agreement between the quantum and the classical calculations seems
to be excellent, except for the nearly degenerate recurrences at
$s/2\pi\approx 1.1$.
}
\end{figure}

We have calculated 3181 eigenvalues $w_n<80$ of the scaling parameter and
the diagonal matrix elements of the two operators $\hat A=1/(2r{\bf p}^2)$
and $\hat B=r{\bf p}^2$ for the hydrogen atom in a magnetic field at
constant scaled energy $\tilde E=-0.1$.
The distributions of the matrix elements are presented in Fig.\ \ref{fig1}.
The matrix elements are distributed randomly around the mean values without
showing any regular pattern, as is typical of systems with chaotic
dynamics.
The quantum mechanical weighted density of states
\begin{equation}
 \rho^{(A)}(w) = \sum_n \langle n|\hat A|n\rangle \delta(w-w_n)
\label{rho_A_qm_scaled:eq}
\end{equation}
can now be analyzed with the harmonic inversion technique to obtain the
scaled actions $s_p$ and the amplitudes $d^{(A)}_p$ (see Eq.\ \ref{semicl})
of the classical periodic orbits.
As can be seen from Eq.\ \ref{rho_A_scaled:eq} the periodic orbit amplitudes
\newpage
\phantom{}
\begin{equation}
 d^{(A)}_p = A_p d_p
\label{d_Ap:eq}
\end{equation}
are given as the product of the amplitudes, $d_p$ of Gutzwiller's original 
trace formula, and the classical periodic orbit means $A_p$ in 
Eq.\ \ref{Ap_s:eq}.
For the graphical presentation of the results it is therefore convenient
to divide the quantum amplitudes $d^{(A)}_p$ obtained by the 
harmonic inversion of the spectra by the amplitudes, $d_p$ of
Gutzwiller's trace formula.
The periodic orbit quantities $A_p$ obtained in this way from the quantum 
spectra at scaled energy $\tilde E=-0.1$ are presented in Fig.\ \ref{fig2} 
for three different operators, viz.\ (a) the identity $\hat I$, and the 
operators (b) $\hat A=1/(2r{\bf p}^2)$ and (c) $\hat B=r{\bf p}^2$.
The solid lines and crosses in Fig.\ \ref{fig2} mark the periodic orbit 
means obtained by the harmonic inversion of the quantum spectra.
For comparison the dashed lines and squares present the periodic orbit 
means of the observable obtained classically via Eq.\ \ref{Ap_s:eq}.
For the identity the classical periodic orbit averages 
(squares in Fig.\ \ref{fig2}a) are exactly equal to one.
This is in excellent agreement with the harmonic inversion analysis of the
quantum mechanical density of states (crosses in Fig.\ \ref{fig2}a), 
despite the two weakly separated periodic orbit contributions around 
$s/2\pi\approx 1.1$.
For the two operators $\hat A=1/(2r{\bf p}^2)$ in Fig.\ \ref{fig2}b and 
$\hat B=r{\bf p}^2$ in Fig.\ \ref{fig2}c the agreement between the 
periodic orbit means obtained by harmonic inversion of the quantum 
spectra and classically by Eq.\ \ref{Ap_s:eq} is of similar high accuracy as 
\newpage
\phantom{}
\noindent
for the identity in Fig.\ \ref{fig2}a.
The results presented in Fig.\ \ref{fig2} demonstrate that harmonic inversion
of quantum spectra \cite{Mai97b} is indeed a powerful tool for the high 
precision check of semiclassical theories.
Fig.\ \ref{fig2} provides an excellent numerical verification, by way of 
example of the hydrogen atom in a magnetic field and the chosen set of 
operators, of the validity of the semiclassical trace formula 
(\ref{rho_A_scaled:eq}) for chaotic systems.

\subsubsection{Regular regime}
In the same way as described above we have checked the validity of the
extended Berry-Tabor formula (\ref{BT_A_scaled:eq}) for integrable systems.
As a physical system we again choose the hydrogen atom in a magnetic field,
but at low scaled energy $\tilde E=-0.5$, where the classical phase space
is regular.
We have calculated 5640 eigenvalues $w_n<160$ of the scaling parameter and
the diagonal matrix elements of the two operators $\hat A=1/(2r{\bf p}^2)$
and $\hat B=r{\bf p}^2$.
The weighted density of states (\ref{rho_A_qm_scaled:eq}) for the identity,
and the operators $\hat A=1/(2r{\bf p}^2)$ and $\hat B=r{\bf p}^2$ have been
analyzed in the same way as explained above.
The results obtained for the regular system at scaled energy $\tilde E=-0.5$ 
resemble those of Fig.\ \ref{fig2} for the chaotic system.
The difference is that the averages of the observables for the resonant tori 
have been extracted from the quantum spectra by application of the 
generalized Berry-Tabor formula (\ref{BT_A_scaled:eq}).
The quantum results perfectly agree with the classical averages which 
illustrates the validity of the generalized Berry-Tabor formula.

\section{Non-trace type formulas}
\label{non-trace:sec}
The generalized semiclassical trace formulas (\ref{rho_A_scaled:eq}) 
and (\ref{BT_A_scaled:eq}) discussed in Sec.\ \ref{trace:sec} allow 
the semiclassical calculation of the diagonal matrix elements of smooth 
operators.
However, it would be desirable to know even more generalized 
expressions for the calculation of {\em products} of matrix elements.
As mentioned in the introduction, such formulas are important, e.g., in 
the semiclassical theory of matrix element fluctuations \cite{Eck95} or 
for the construction of cross-correlated periodic orbit sums \cite{Mai99}.
To study matrix element fluctuations of an operator $\hat A$ the density 
of states can be weighted with the variances 
${\rm Var}_n A\equiv\langle n|\hat A^2|n\rangle-\langle n|\hat A|n\rangle^2$,
i.e.
\begin{eqnarray}
 \rho^{({\rm Var}~A)}(E) &\equiv&
     \sum_n \langle n|\hat A^2|n\rangle \delta(E-E_n) \nonumber \\
 &-& \sum_n \langle n|\hat A|n\rangle^2 \delta(E-E_n) \; .
\label{rho_Var_A_qm:eq}
\end{eqnarray}
The first term in (\ref{rho_Var_A_qm:eq}) can be written as a semiclassical
trace formula (Eqs.\ \ref{rho_A_sc:eq} and \ref{BT_A:eq} for chaotic and
regular systems, respectively) with the observable $A$ replaced with its
square, $A^2$.
However, because of the squares of the matrix elements, the second 
term in (\ref{rho_Var_A_qm:eq}) in general cannot be expressed in a 
straightforward fashion with the help of the Green's operator $\hat G_E^+$ 
as a trace formula.
The trivial exception is when the operator $\hat A$ commutes with the 
Hamiltonian, which means that $A$ is a constant of motion and thus its 
variance vanishes.
Thus the derivation of a semiclassical approximation to the second term 
in (\ref{rho_Var_A_qm:eq}) constitutes a nontrivial problem.

One solution can be obtained by application of periodic orbit sum rules
\cite{Ber85}.
Using smooth approximations of the $\delta$-functions, e.g.\ Gaussians
of width $\epsilon$,
\begin{equation}
 \delta_\epsilon(E) = {1\over\sqrt{2\pi}\epsilon} e^{-E^2/2\epsilon^2} \; ,
\end{equation}
and the relation
\begin{equation}
   \delta_\epsilon^2(E)
 = {1\over 2\sqrt{\pi}\epsilon} \delta_{\epsilon/\sqrt{2}}(E)
\end{equation}
the second term in (\ref{rho_Var_A_qm:eq}) can formally be written as
the square of the density of states weighted with the diagonal matrix 
elements \cite{Ber85,Eck95}, viz.
\begin{eqnarray}
 & &   \sum_n \langle n|\hat A|n\rangle^2
       \delta_{\epsilon/\sqrt{2}} (E-E_n) \nonumber \\
 &=& 2\sqrt{\pi}\epsilon\sum_n \langle n|\hat A|n\rangle^2
       \delta_\epsilon^2 (E-E_n) \nonumber \\
 &=& 2\sqrt{\pi}\epsilon\left[\sum_n \langle n|\hat A|n\rangle
       \delta_\epsilon (E-E_n)\right]^2 \; .
\label{Berry_trick:eq}
\end{eqnarray}
The width $\epsilon$ in (\ref{Berry_trick:eq}) must be chosen sufficiently
small so that the smoothed $\delta$-functions do not overlap.
However, it should be noted that this condition cannot be fulfilled for
systems with degenerate states.
On the r.h.s.\ of Eq.\ \ref{Berry_trick:eq} the weighted density of states
can now be written as a trace formula and replaced with its semiclassical
approximations (\ref{rho_A_sc:eq}) and (\ref{BT_A:eq}) for chaotic and
regular systems, respectively.
Evaluation of the square of the periodic orbit sum on the r.h.s.\ of
(\ref{Berry_trick:eq}) then finally yields a double sum over all periodic
orbits of the classical system.
Although this result is formally correct, it is very inconvenient for
practical applications for the following reasons.
Firstly, the number of periodic orbits proliferates exponentially in
chaotic systems and the handling of the single periodic orbit sum is 
already nontrivial.
The practical evaluation of the double sum would be even more cumbersome.
Secondly, the width $\epsilon$ in (\ref{Berry_trick:eq}) is a free
parameter.
Although the results should not depend on the width if $\epsilon$ is chosen
sufficiently small, the appropriate choice may render numerical calculations
extremely expensive.
Thirdly, and most importantly, the r.h.s.\ of Eq.\ \ref{Berry_trick:eq} does 
not coincide with the ``simple'' trace formulas in those special cases, 
where the operator $\hat A$ commutes with the Hamiltonian.
Even for the simplest operator, the identity $\hat A=\hat I$, 
we end up with the nontrivial periodic orbit sum rule of Ref.\ \cite{Ber85} 
instead of Gutzwiller's trace formula for the density of states.
Especially the third point indicates that the procedure described above
might not be the simplest way to construct a semiclassical approximation 
to non-trace type formulas such as Eq.\ \ref{rho_Var_A_qm:eq}.
It is the main objective of this section to present a semiclassical 
approximation to non-trace type weighted densities of states.
Our semiclassical expressions agree with the well established ``simple'' 
semiclassical trace formulas when the weighted density of states can be 
written, for at least one of the operators commuting with the Hamiltonian, 
as a quantum mechanical trace formula.

Starting from a more general equation than (\ref{rho_Var_A_qm:eq})
we study the density of states
\begin{equation}
 \rho^{(A,B)}(E) \equiv \sum_n \langle n|\hat A|n\rangle
                 \langle n|\hat B|n\rangle \delta(E-E_n) \; ,
\label{rho_AB_qm:eq}
\end{equation}
weighted with the product of the diagonal matrix elements of two smooth 
operators $\hat A$ and $\hat B$.
Eq.\ \ref{rho_AB_qm:eq} is the starting point to construct a quantum
mechanical cross-correlation function from a set of operators in Ref.\ 
\cite{Mai99}.
The variance of matrix elements (Eq.\ \ref{rho_Var_A_qm:eq}) is obtained 
by setting $\hat B=\hat A$.
The weighted density of states (\ref{rho_AB_qm:eq}) can only be written
as a trace formula, 
$\rho^{(A,B)}(E)=(-1/\pi)\,{\rm Im~tr}\,\{\hat A\hat G_E^+\hat B\}$
if either $\hat A$ or $\hat B$ commutes with the Hamiltonian.
As discussed in Sec.\ \ref{trace:sec} (see Eq.\ \ref{d_Ap:eq})
the semiclassical expressions for the weighted densities of states differ
from Gutzwiller's trace formula and the Berry-Tabor formula in the
following way.
The periodic orbit amplitudes are multiplied with the classical periodic
orbit (or torus) averages of the observable $A$.
We now assume that this ansatz is still valid for the non-trace type weighted
density of states (\ref{rho_AB_qm:eq}), i.e., its semiclassical analogue
has the same functional form as Gutzwiller's periodic orbit sum but with
periodic orbit amplitudes $d_p$ multiplied with the classical averages 
$A_p$ and $B_p$ of both observables $A$ and $B$,
\begin{equation}
 d^{(A,B)}_p = A_p B_p d_p \; ,
\label{d_ABp:eq}
\end{equation}
with $A_p$ and $B_p$ given by Eqs.\ \ref{Ap_t:eq} and \ref{Ap_reg:eq}
for chaotic and regular systems, respectively.
As can easily be seen, this ansatz has the property that the trace formulas 
(\ref{rho_A_sc:eq}) and (\ref{BT_A:eq}) are recovered if one of the 
operators is chosen to be the identity or one of the operators commutes 
with the Hamiltonian.
However, the general validity of this ansatz is not at all obvious, and will 
be checked numerically by the high resolution analysis of quantum spectra 
in the following.
With the ansatz (\ref{d_ABp:eq}) for the periodic orbit amplitudes the
semiclassical analogue to the non-trace type formula (\ref{rho_AB_qm:eq}) 
reads
\begin{eqnarray}
\label{rho_AB_sc:eq}
 & & \rho^{(A,B)}(E) = \rho^{(A,B)}_0(E) \\
 &+& {1\over\pi\hbar}\, {\rm Re}\, \sum_p A_p B_p \sum_{r=1}^\infty
 {T_p\over\sqrt{|\det(M_p^r-I)|}}\,e^{i[S_p(E)/\hbar-{\pi\over 2}\mu_p]r} \; ,
 \nonumber
\end{eqnarray}
for systems with underlying chaotic classical dynamics, and
\begin{eqnarray}
\label{BT_AB:eq}
 & & \rho^{(A,B)}(E) = \rho^{(A,B)}_0(E) \\
 &+& {1\over\pi\hbar^{3/2}}\, {\rm Re}\, \sum_{\bf M} A_{\bf M} B_{\bf M}
 {T_{\bf M}\over M_2^{3/2}|g_E''|^{1/2}} \,
 e^{i[S_{\bf M}(E)/\hbar-{\pi\over 2}\eta_{\bf M}-{\pi\over 4}]}\, ,
 \nonumber
\end{eqnarray}
for integrable systems.
Eqs.\ \ref{rho_AB_sc:eq} and \ref{BT_AB:eq} are the central propositions
of this paper, and generalize the semiclassical trace formulas
(\ref{rho_A_sc:eq}) and (\ref{BT_A:eq}) to the non-trace type weighted 
density of states (\ref{rho_AB_qm:eq}).
The nontrivial statement is that weighting the quantum 
mechanical density of states with the product of diagonal matrix elements
of smooth operators is equivalent, on the semiclassical level, to weighting 
the periodic orbit contributions in the periodic orbit sum with the product 
of the averages of the corresponding classical observables.

In analogy to the discussion of scaling properties in Sec.\ \ref{trace:sec},
Eqs.\ \ref{rho_AB_sc:eq} and \ref{BT_AB:eq} can be reformulated for scaling 
systems, viz.
\begin{eqnarray}
\label{rho_AB_scaled:eq}
 & & \rho^{(A,B)}(w)
  =  \rho^{(A,B)}_0(w) \\
 &+& {1\over\pi\hbar}\, {\rm Re}\, \sum_p A_p B_p \sum_{r=1}^\infty
     {s_p\over\sqrt{|\det(M_p^r-I)|}}\, e^{i[s_p w-{\pi\over 2}\mu_p]r}
 \nonumber
\end{eqnarray}
for chaotic systems, and
\begin{eqnarray}
\label{BT_AB_scaled:eq}
 & & \rho^{(A,B)}(w) = \rho^{(A,B)}_0(w) \\
 &+& {1\over\pi\hbar^{3/2}}\, {\rm Re}\, \sum_{\bf M} A_{\bf M} B_{\bf M}
 {s_{\bf M}\over M_2^{3/2}|g_E''|^{1/2}} \,
 e^{i[s_{\bf M}w-{\pi\over 2}\eta_{\bf M}-{\pi\over 4}]}
 \nonumber
\end{eqnarray}
for two-dimensional systems with regular dynamics.
For scaling systems the classical periodic orbit averages $A_p$ and $B_p$
in (\ref{rho_AB_scaled:eq}) must be calculated with respect to the 
classical action instead of time as defined in Eq.\ \ref{Ap_s:eq}.

In the following we will provide convincing numerical evidence for the 
validity of the semiclassical non-trace type formulas by the high precision 
analysis (harmonic inversion) of quantum spectra of two different systems, 
viz.\ the hydrogen atom in a magnetic field and the circle billiard.
A rigorous mathematical proof of the expressions given above is still
lacking and constitutes a challenge for the further development of 
semiclassical theories.

\subsection{Hydrogen atom in a magnetic field}
\begin{figure}
\vspace{9.5cm}
\includegraphics{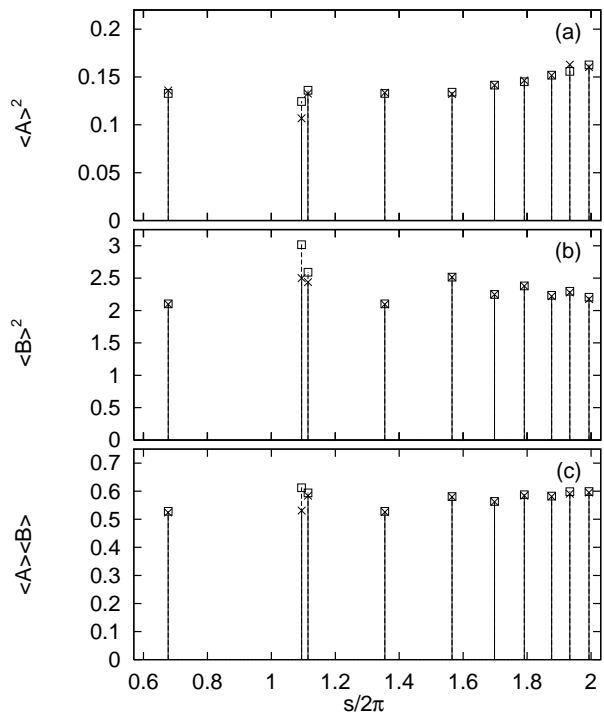}
\caption{\label{fig5} 
Products of periodic orbit means of the two observables $A=1/(2r{\bf p}^2)$ 
and $B = r{\bf p}^2$ for the hydrogen atom in a magnetic field at scaled 
energy $\tilde E=-0.1$ as functions of the scaled action.
Solid lines and crosses: Results of the harmonic inversion of the non-trace
type weighted densities of states.
Dashed lines and squares: Results obtained by classical calculations.
As in Fig.\ \ref{fig2}, the agreement between the quantum and the classical 
calculations seems to be excellent, except for the nearly degenerate 
recurrences at $s/2\pi\approx 1.1$.
}
\end{figure}

To demonstrate the validity of the semiclassical non-trace type formulas,
Eqs.\ \ref{rho_AB_scaled:eq} and \ref{BT_AB_scaled:eq}, we use the same 
system and set of operators as in Sec.\ \ref{trace:sec}, viz.\ the hydrogen 
atom in a magnetic field at scaled energies $\tilde E=-0.1$ and 
$\tilde E=-0.5$ in the chaotic and near-integrable regime, respectively, 
and the operators $\hat A=1/(2r{\bf p}^2)$ and $\hat B=r{\bf p}^2$.
With the quantum mechanical eigenvalues and diagonal matrix elements at hand,
we construct the weighted densities of states (see Eq.\ \ref{rho_AB_qm:eq})
(a) $\rho^{(A,A)}(w)$, (b) $\rho^{(B,B)}(w)$, and (c) $\rho^{(A,B)}(w)$.
These spectra are analyzed with the harmonic inversion technique as described
in Sec.\ \ref{trace:sec}.
The analysis provides the scaled action $s_p$ of the periodic orbits and the 
periodic orbit amplitudes $d^{(A,A)}_p$ ($d^{(B,B)}_p$ and $d^{(A,B)}_p$).
The amplitudes of the weighted densities of states are divided by the 
amplitudes of the unweighted densities of states to obtain, according to 
Eq.\ \ref{d_ABp:eq}, the products of the periodic orbit means $A_p^2$ 
($B_p^2$ and $A_pB_p$).
These values are presented as solid lines and crosses in Fig.\ \ref{fig5}
for the spectra in the chaotic regime at scaled energy $\tilde E=-0.1$ and
in Fig.\ \ref{fig6} for the spectra at scaled energy $\tilde E=-0.5$ in the
near-integrable regime.
For comparison, the squares mark the products of the periodic orbit means
obtained from the classical calculations.
As in Fig.\ \ref{fig2} for the high precision check of the
semiclassical trace formula (\ref{rho_A_scaled:eq}), the agreement between 
the quantum and classical calculations is found to be very good, which 
strongly supports the validity of the semiclassical non-trace type expressions.
Note that the somewhat larger deviations between the crosses and squares
for the nearly degenerate recurrencies at $s_p/2\pi\approx 1.1$ in Fig.\ 
\ref{fig5} have also been observed in Fig.\ \ref{fig2} for the semiclassical 
trace formulas, i.e., the deviation does not indicate any failure of the 
non-trace type formula (\ref{rho_AB_scaled:eq}).
\newpage
\phantom{}

\subsection{Circle billiard}
We now investigate the validity of the semiclassical non-trace type formula
(\ref{BT_AB_scaled:eq}) on a second system, viz.\ the integrable circle 
billiard.
This system also serves as a model example in Ref.\ \cite{Mai99} to 
construct a semiclassical cross-correlated periodic orbit sum for a given 
set of smooth observables, and to calculate semiclassical spectra and 
diagonal matrix elements by harmonic inversion of the cross-correlation 
function.
As is well known, Schr\"odinger's equation for the circle billiard with 
radius $R$ can be separated in polar coordinates $(r,\varphi)$, and the 
wave functions can be expressed in terms of Bessel functions,
\begin{equation}
   \psi_{nm}(r,\varphi)
 = {\cal N}_{nm} J_{|m|}(k_{nm}r)e^{im\varphi} \; ,
\label{Bessel:eq}
\end{equation}
with the ${\cal N}_{nm}$ being normalization constants, $m$ the angular 
momentum quantum number, and $k_{nm}=\sqrt{2ME_{nm}}/\hbar$ the quantized wave
numbers obtained as the $n$th zero of Bessel functions, $J_{|m|}(k_{nm}R)=0$.
In the following we choose radius $R=1$.
We calculated 
\newpage
\phantom{}
\begin{figure}
\vspace{9.2cm}
\includegraphics{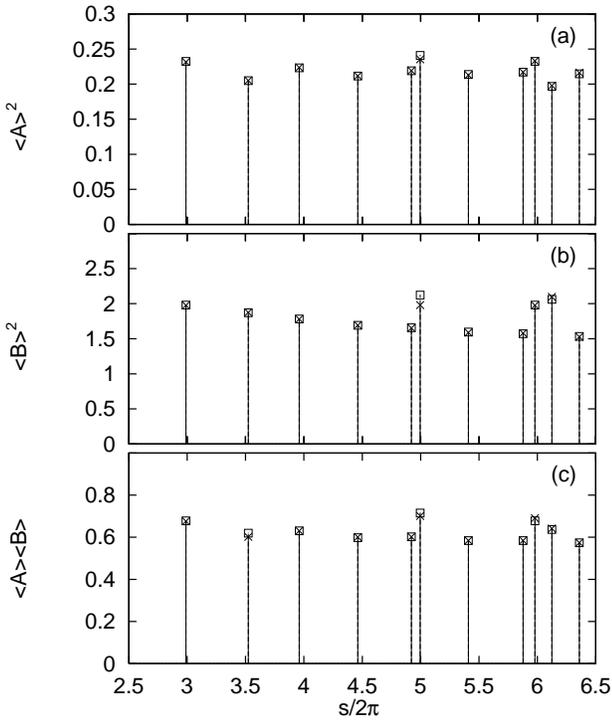}
\caption{\label{fig6} 
Same as Fig.\ 5 but at scaled energy $\tilde E=-0.5$.
}
\end{figure}
\noindent
31208 eigenvalues $k_{nm}<500$, and the diagonal matrix
elements of the operators $r$ and $r^2$.
The quantum spectra of (a) the unweighted density of states $\rho(k)$, 
(the wave number $k$ is the scaling parameter, $w=k$ for billiard systems 
\cite{Mai99}) and the density of states weighted with the matrix element 
expressions
(b) $\langle\psi_{nm}|r^2|\psi_{nm}\rangle$,
(c) $\langle\psi_{nm}|r|\psi_{nm}\rangle^2$, and
(d) the variance $\langle{\rm Var}~r\rangle_{nm}\equiv
\langle\psi_{nm}|r^2|\psi_{nm}\rangle-\langle\psi_{nm}|r|\psi_{nm}\rangle^2$
have been analyzed with the harmonic inversion method.
The amplitudes obtained, divided by the amplitudes of the Berry-Tabor formula,
are presented as solid lines and crosses in Fig.\ \ref{fig7}, and the
corresponding classical averages are drawn as squares for comparison.
As can be seen, the agreement is perfect, not only for the identity and the
periodic orbit means of the observable $r^2$ in Fig.\ \ref{fig7}a and 
\ref{fig7}b, verifying the Berry-Tabor formula and its extension 
(\ref{BT_A_scaled:eq}), but also for the squares of the periodic orbit means 
of $r$ and the variance of this observable in Fig.\ \ref{fig7}c and 
\ref{fig7}d, where the agreement demonstrates the validity of the non-trace
type equation (\ref{BT_AB_scaled:eq}) for the circle billiard with 
$\hat A=\hat B=r$.
The squares in Fig.\ \ref{fig7}d mark the classical variances of 
the observable $r$ on the various resonant tori.
Our conjecture therefore provides a basic formula for semiclassical matrix
element fluctuations, since it directly relates the quantum variances 
${\rm Var}_n A\equiv\langle n|\hat A^2|n\rangle-\langle n|\hat A|n\rangle^2$
of a smooth operator $\hat A$ to the classical variances
${\rm Var}_p A\equiv \langle A^2\rangle_p-\langle A\rangle_p^2$
of the observable $A$ taken along the periodic orbits or resonant tori.

The perfect agreement between the quantum and classical results for the 
circle billiard in Fig.\ \ref{fig7} compared to
\newpage
\phantom{}
\begin{figure}
\vspace{10.5cm}
\includegraphics{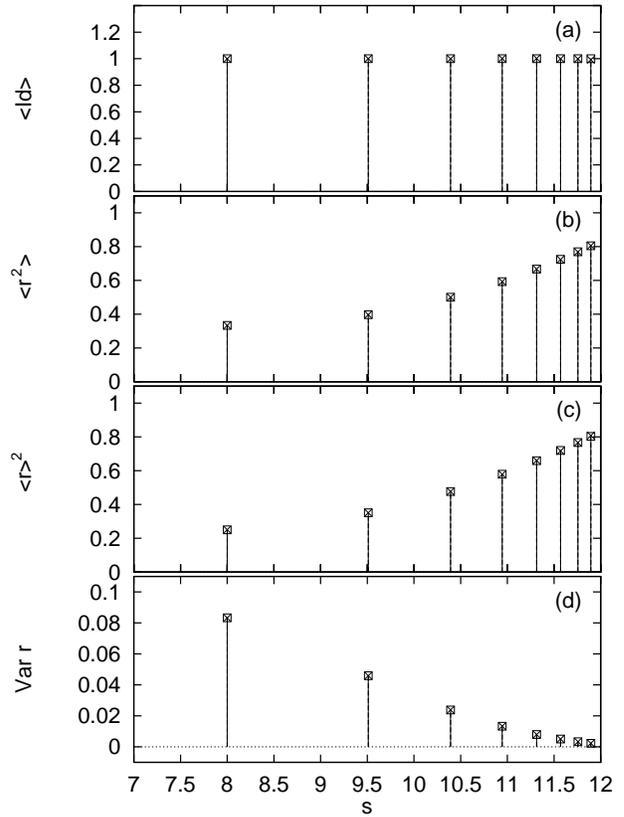}
\caption{\label{fig7} 
Classical averages on rational tori of (a) the identity, 
(b) $\langle r^2\rangle_p$, (c) $\langle r\rangle_p^2$, and (d) the variances
${\rm Var}_p r\equiv\langle r^2\rangle_p-\langle r\rangle_p^2$
for the circle billiard with radius $R=1$ as functions of the scaled action.
Solid lines and crosses: Results of the harmonic inversion of quantum spectra.
Dashed lines and squares: Periodic orbit means obtained by classical
calculations.
}
\end{figure}
\noindent
the very good but not
absolutely perfect results for the hydrogen atom in a magnetic field 
in Figs.\ \ref{fig5} and \ref{fig6} may be explained by the different 
number of quantum states used in the harmonic inversion analysis.
For the circle billiard we have calculated more than 30000 states, which 
is by about a factor of 10 (5.5) times more quantum states than for 
the hydrogen atom in a magnetic field at scaled energy $\tilde E=-0.1$ 
($\tilde E=-0.5$).

\section{Conclusion and outlook}
\label{conclusion}
We have extended semiclassical trace formulas for the density of states
of regular and chaotic systems, or the density of states weighted with
the diagonal matrix elements of smooth operators, to the more general
class of {\em non-trace} type equations, where the density of states
is weighted with the diagonal matrix elements of two operators $\hat A$
and $\hat B$, i.e., $\rho^{(A,B)}(E)
=\sum_n\langle n|\hat A|n\rangle\langle n|\hat B|n\rangle\delta(E-E_n)$.
By the high resolution analysis (harmonic inversion) of the 
quantum spectra of two different systems, viz.\ the hydrogen atom in a 
magnetic field and the circle billiard, we have given numerical evidence
that weighting the quantum mechanical density of states with the product 
of the diagonal matrix elements 
$\langle n|\hat A|n\rangle\langle n|\hat B|n\rangle$
is equivalent, on the semiclassical level, to weighting the periodic orbit 
contributions in the periodic orbit sum with the product of the averages 
of the corresponding classical observables,
$\langle A\rangle_p\langle B\rangle_p$, where the means are taken along 
the periodic orbits or resonant tori for chaotic and regular systems,
respectively.
However, a rigorous mathematical derivation of semiclassical non-trace 
type formulas appears nontrivial, and would be a challenging task for the 
further development of semiclassical theories.

There are several useful and important applications of semiclassical 
non-trace type formulas.
For example, it enables the semiclassical approach to matrix element
fluctuations. The variances 
${\rm Var}_n A\equiv\langle n|\hat A^2|n\rangle-\langle n|\hat A|n\rangle^2$
of the diagonal matrix elements of a smooth operator $\hat A$ are expressed 
in terms of the variances
${\rm Var}_p A\equiv\langle A^2\rangle_p-\langle A\rangle_p^2$
of the classical observable $A$ taken along the periodic orbits or
resonant tori.
Non-trace type formulas also provide the semiclassical approximation to 
cross-correlated weighted density of states,
$\rho_{\alpha\alpha'}(E)=\sum_n\langle n|\hat A_\alpha|n\rangle
\langle n|\hat A_{\alpha'}|n\rangle\delta(E-E_n)$
with a set of smooth operators $\hat A_\alpha$, $\alpha=1,\dots,D$.
The additional classical information obtained from the set of classical
observables can be used to significantly improve the convergence properties 
of semiclassical quantization methods \cite{Mai99}.

In this paper we have investigated non-trace type expressions for products 
of two diagonal matrix elements.
These products have been chosen because of the important applications
to semiclassical matrix element fluctuations, i.e., the calculation of
variances of matrix elements and to the semiclassical quantization method
in Ref.\ \cite{Mai99}.
However, our conjecture is not restricted to products of two matrix elements.
For example, it appears straightforward to generalize Eqs.\ \ref{rho_AB_sc:eq}
and \ref{BT_AB:eq} to products of more than two matrix elements and classical 
periodic orbit means.
The most general case of non-trace type equations would be the analysis of
functions $f(A_{nn},B_{nn},C_{nn},\dots)$ of one or more diagonal matrix
elements, i.e.,
$\rho^{(f)}(E)=\sum_nf(A_{nn},B_{nn},C_{nn},\dots)\delta(E-E_n)$,
which should be obtained semiclassically by multiplying the
periodic orbit amplitudes of Gutzwiller's trace formula or the Berry-Tabor
formula with the function 
$f(\langle A\rangle_p,\langle B\rangle_p,\langle C\rangle_p,\dots)$
of the periodic orbit means of the corresponding classical observables.
Certainly the operators and the function $f$ must be smooth.
Clearly, further investigations will be necessary to verify that conjecture
and to specify the smoothness conditions on operators and functions.

In conclusion, the analysis of non-trace type equations will provide a
valuable instrument for extending the relation between quantum mechanical 
matrix elements on the one side and the periodic orbit means of classical 
observables on the other.

\acknowledgements
We acknowledge stimulating discussions with J.\ Keating.
This work was supported in part by the Son\-der\-for\-schungs\-be\-reich 
No.\ 237 of the Deutsche For\-schungs\-ge\-mein\-schaft.
J.M.\ is grateful to Deutsche For\-schungs\-ge\-mein\-schaft for a
Habilitandenstipendium (Grant No.\ Ma 1639/3).\\[-3ex]

\end{document}